# A QoS ontology-based Component selection

Lamia Yessad[1] and Zizette Boufaida[2]

[1] University 20 Août 1955 of Skikda, Algeria
`yessad_lamia@yahoo.fr`
[2] Lire Laboratory, University Mentouri of Constantine, Algeria
`zboufaida@gmail.com`

## ABSTRACT

*In the component-based software development, the selection step is very important. It consists of searching and selecting appropriate software components from a set of candidate components in order to satisfy the developer-specific requirements. In the selection process, both functional and non-functional requirements are generally considered. In this paper, we focus only on the QoS, a subset of non-functional characteristics, in order to determine the best components for selection. The component selection based on the QoS is a hard task due to the QoS descriptions heterogeneity. Thus, we propose a QoS ontology which provides a formal, a common and an explicit description of the software components QoS. We use this ontology in order to semantically select relevant components based on the QoS specified by the developer. Our selection process is performed in two steps: (1) a QoS matching process that uses the relations between QoS concepts to pre-select candidate components. Each candidate component is matched against the developer's request and (2) a component ranking process that uses the QoS values to determine the best components for selection from the pre-selected components. The algorithms of QoS matching and component ranking are then presented and experimented in the domain of multimedia components.*

## 1. INTRODUCTION

The paradigm of Component-Based Software Engineering (CBSE) aims to develop software by assembling and deploying reusable units, called software components. In order to be queried and reused in software, the components must have a shared description. Several standardised description models [1, 2] were proposed and are focused on the functional characteristics of components. Unfortunately, they don't consider the QoS aspects which are decisive in the selection process. However, there are CBSE researches [3, 4, 5] that address the component selection problem based on specific QoS models. The shortcomings of these approaches are mainly threefold: (1) The QoS models are not semantic (2) once implemented they are hardly extensible and (3) the implementations are plateforms-dependent.

In our work, we focus on the QoS characteristics in order to select the most relevant components for the developer's requirements. The QoS is a subset of non-functional characteristics of components that have an impact on the final users' expectations. Components with same functionalities may have quite different QoS. Thus, QoS is important in the component selection process. This latter is a hard task due to the QoS descriptions heterogeneity. The QoS is a multivalued concept composed of several QoS characteristics which are for instance:

- Reliability: is the ability of a system or component to perform the required functions under normal conditions in a period of time;
- Performance: is the degree to which a system or component accomplishes its functions under constraints. Thus, performance refers to Throughput (the number of requests served in a given time period), Response Time (the delay from the request to getting a response from a component interface) or Latency (the time between client request and the start of the server response);
- Robustness: is the ability of a system or a component to perform the required functions under abnormal conditions;
- Cost: the amount of money for the component acquisition.

In CBSE field, one QoS characteristic may be described by multiple terms and the same term may design multiple QoS characteristics. The majority of CBSE works neglect the semantic aspect of QoS. Whereas, several SOA-based works [6, 7] use ontologies to deal with the QoS interoperability problem. The benefits of the QoS concepts modelling by ontologies are:

- Ontology provides a shared and common description for the QoS and thus, reduces the semantic heterogeneity between component descriptions;
- It provides reasoning mechanisms on QoS concepts;
- It is extensible, i.e., new QoS concepts and properties, can be added.

In the scope of this paper, we propose a QoS ontology in order to achieve two main objectives (*cf.* Figure 1):

- Describing the QoS of software components;
- Selecting the relevant components for developer's QoS requirements.

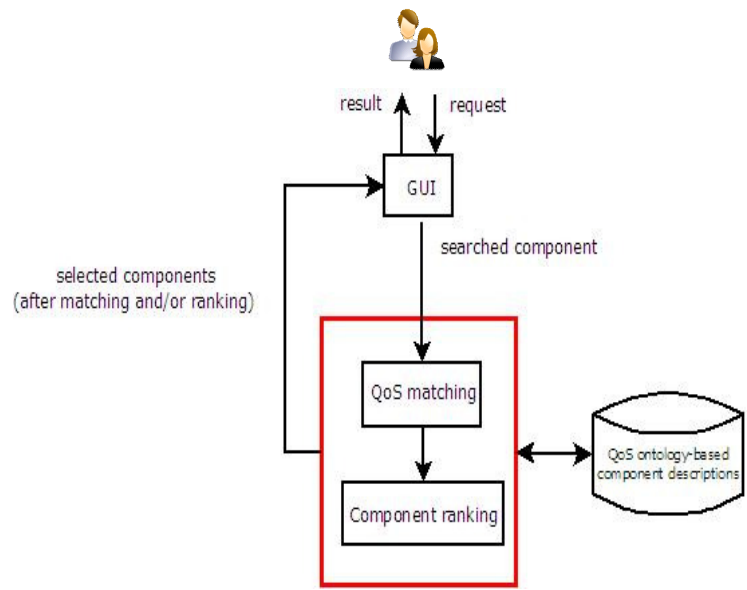

Figure 1.  A QoS ontology-based component selection architecture

The software component is the central concept in the QoS ontology. It is defined as a set of provided and required interfaces. Furthermore, the QoS of a provided interface depends on the QoS of required interfaces. For instance, in [8], the reliability of a component has to be modelled as a function having the reliability of external services as parameters.

We construct an ontology by drawing classes and properties from existing QoS ontologies [9, 10]. We use the concept "QoSMetric" and the concept "QoSfunction" with an adapted measurement process. Particularly, we are interested to the component selection process which is performed in two steps:

- The QoS matching process which compares the QoS of components' interfaces;
- The component ranking process which ranks components according to their matching results and a dissimilarity measure (*cf.* section 4.2).

The paper is organised as follows: In section 2, we first discuss some related works both in CBSE and SOA fields. In section 3, we present the definition of the QoS ontology that we propose. In section 4, we describe, based on our QoS ontology, a matching algorithm and a ranking algorithm, that are used in order to select the relevant components for the developer's QoS requirements. Finally, we present some experiment results which measure the performance of our approach.

## 2. RELATED WORKS

### 2.1. QoS in Component-oriented approaches

We analysed several QoS Specification languages and models [3, 4, 11] and we identified from them the concepts that are adapted to our research goal. We use these concepts in our ontology. QML [12], CQML [13] and CQML+ [14] are specification languages for component QoS constraints. These works describe categories of constraints like reliability, security and performance. A category has one or more dimensions. For instance, the dimensions such as MTTF (Mean Time To Failure), MTTR (Mean Time To Repair) and MTBF (Mean Time Between Failures) refer to the reliability. Each dimension has a metric that can be either numeric for the quantitative characteristics or ordinal for the qualitative characteristics.

The link between the QoS and the functional characteristics is represented by a construction called QoSProfile. CQML proposes matching rules for simple or compound QoS profiles [13]:

- If C and D are simple profiles. C is conform to D if C provides at least the same and expects at most the same properties as D;
- A simple profile C is conform to a compound profile D if C is conform to each simple profile that is part of D;
- If C and D are compound profiles, C is conform to D if each simple profile in C is conform to a simple profile in D.

However, [13] doesn't provide any comparison between the values of properties. So, these approaches lack of semantic expressivity and ranking mechanisms [15].

### 2.2. QoS in service-oriented approaches

Most existing QoS ontologies refer to three ontology levels: The ontology that allows the association of QoS with the service profile (functional aspect), the ontology that defines QoS characteristics and QoS metrics, and the ontology that is domain-specific.

In [10], the onQoS ontology contains the upper, middle and low ontologies. The structure of the upper ontology is motivated by the need to provide a link to the OWL-S ontology and the means to specify QoS measurements. The middle ontology defines the standard vocabulary related to QoS such as QoS parameters, QoS metrics and QoS scales. Finally the low ontology defines the concepts, the properties and the constraints of a specific domain. It also makes measurements of different metric types such as derived, internal and external metrics.

In [9], the QoS property definition layer of the DAML-QoS ontology determines both the domain and the range constraints of QoS properties which are used in the QoS matching mechanisms. It also defines atomic and complex metrics.

The OWL-Q ontology [16] is an upper level ontology which allows the association of a QoS specification with an OWL-S service profile. It also provides means to specify composite metrics. However, it is has no support to construct a QoS profile from a set of QoS attributes, i.e. QoS characteristics.

The WSMO-QoS ontology [17] also does not provide any means to specify a QoS profile from a set of QoS characteristics.

The QoSOnt ontology [18] contains several ontologies. The usage domain ontology connects ontologies to a specific type of systems, for instance network system or Web service system. Whereas all these approaches construct ontologies for metrics and allow unit conversions, only some ones [9] [10] employ the ontology reasoning for matching purposes.

In our research, we use the ontology model for describing the QoS characteristics in order to (1) avoid semantic ambiguity between terms and concepts (2) explicitly describe the domain-dependent functions that are specified between QoS metrics and (3) use the ontology reasoning mechanisms for matching purposes.

## 3. PROPOSED QoS ONTOLOGY

We construct a QoS ontology for the selection process of software components (*cf*. Figure 2). It is organized in classes and properties hierarchies, and is serialized to OWL[1] (Web Ontology Language).

A software component has interfaces that describe the functional characteristics (services) of the component. We consider two types of component interfaces: the provided interfaces and the required interfaces. In order to describe the QoS characteristics, we construct a QoSProfile class. Each component interface (provided or required) is related to an instance of the QoSProfile class.

The object property hasCharacteristic relates a QoSProfile instance to a QoSCharacteristic instance. We distinguish between two kinds of QoS characteristics:

- Service characteristics that represent high level characteristics related, for instance, to video and audio services such as frame rate, video resolution and audio sampling.
- Resource characteristics that represent resources related to the operating system and/or the underlying network such as CPU rate, network throughput and latency.

We consider that *QoSCharacteristic* has one or more metrics, instances of the *QoSMetric* class that measure the quality level of a characteristic. We describe a QoS metric by a minValue, a maxValue, a processMeasurement and a direction. These datatype properties of the *QoSMetric* class have the following significations:

---

[1] http://www.w3.org/2004/OWL/

- minValue: When the QoS metric is not an exact value, we have to specify an interval whose the lower bound is minValue;
- maxValue: it is the upper bound of the interval of the QoS metric values. When maxValue is equal to minValue, the QoS metric has a fixed value.

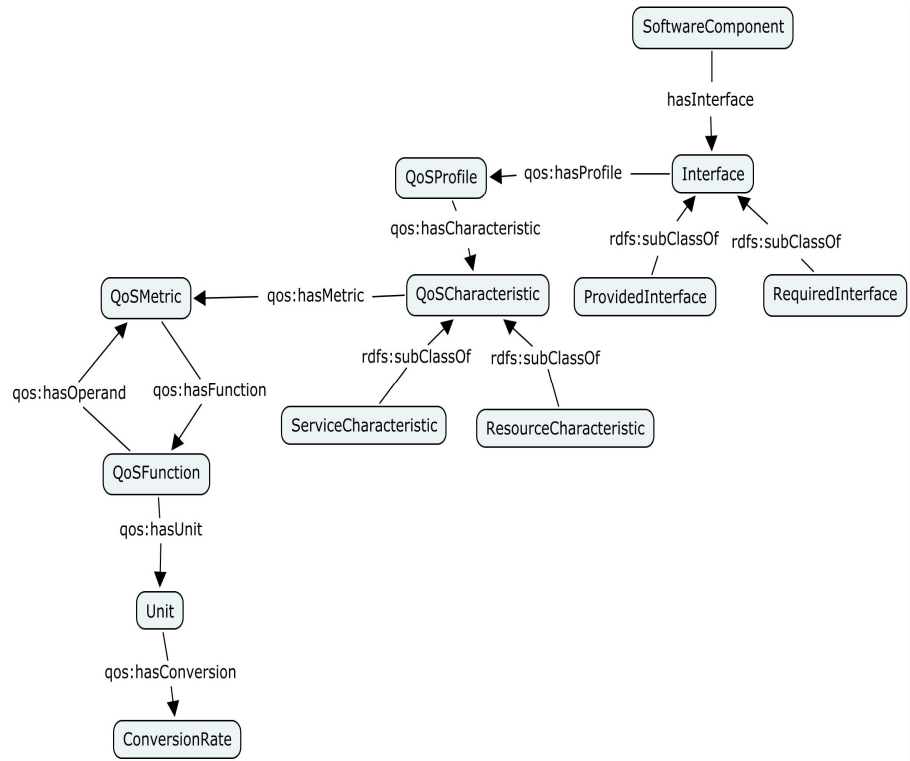

Figure 2. An excerpt of the QoS ontology

- processMeasurement: it links the *QoSMetric* class to the *QoSFunction* class when the value of a QoS metric is not given;
- direction: it has two values: increasing and decreasing ones. An increasing value means that the quality is best when the metric value is highest (for example the frame rate). Inversely, a decreasing direction means that the quality is best when the metric value is lowest (for instance, the response time of an operation).

In this paper, a metric value may be given or calculated using a predictable model. The *QoSFunction* class contains the measurement process and defines the different operands which are *QoSMetric* instances. Thus, *hasOperand* relates a *QoSMetricFunction* instance to a *QoSMetric* instance. A QoS function is parameterised by a vector of metrics values. A measurement process is performed to determine all the QoS metrics values of the provided interfaces.

Finally, a QoS metric has a *Unit*. Several units are associated to the same quality metrics. Thus, it is necessary to consider the conversion rate (*ConversionRate* class) between units. For example, the operation response time can be measured in s (second), ms (millisecond) or μs (microsecond) and a conversion process has to indicate, for instance, that 1 s equals $10^3$ ms or $10^6$ μs.

```
<owl:DatatypeProperty rdf:ID="value">
    <rdfs:domain rdf:resource="#QoSMetric"/>
    <rdfs:range rdf:resource="&xsd;#integer"/>
</owl:DatatypeProperty>
<owl:DatatypeProperty rdf:ID="direction">
    <rdfs:domain rdf:resource="#QoSMetric"/>
    <rdfs:range rdf:resource="&xsd;#boolean"/>
</owl:DatatypeProperty>
<owl:ObjectProperty rdf:ID="hasFunction">
    <rdfs:domain rdf:resource="#QoSMetric"/>
    <rdfs:range rdf:resource="QoSFunction"/>
</owl:ObjectProperty>
<owl:ObjectProperty rdf:ID="hasOperand">
    <rdfs:domain rdf:resource="#QoSFunction"/>
    <rdfs:range rdf:resource="#QoSMetric"/>
</owl:ObjectProperty>
<owl:ObjectProperty rdf:ID="hasUnit">
    <rdfs:domain rdf:resource="#QoSMetric"/>
    <rdfs:range rdf:resource="#Unit"/>
</owl:ObjectProperty>
```

Figure 3.  Examples of QoS metric properties

The ontology described above is application-independent. It may be extended to a domain-specific ontology such as a multimedia QoS ontology. Therefore, application-dependant concepts may reuse the classes and the properties of our application-independent ontology.

For instance, we use the QoS ontology in order to describe the QoS of a camera video component which has two provided interfaces (VideoStream and CameraControl types) and a required interface (DVFormat type) [15]. The QoS of this component is specified as follows:

• The video camera has to send at least 25 fps (frames per second);
• It has to start at latest 10 milliseconds after the "start" method call;
• It requires a reliable DV component at 99 percent.

This QoS specification is serialized to OWL by reusing classes and properties of our QoS ontology.

## 4. MATCHING AND RANKING PROCESSES

## 4.1. QoS matching process

The specification matching is "a process of determining if two software components are related" [19]. In our work, we consider that the functional matching is already verified and we focus on the QoS matching (using the QoS ontology). In order to be practical, we make some assumptions to define the framework of our research:

• The software component is defined as a set of provided and required interfaces. An interface is an abstraction of a component and represents its visible part;
• The QoS features of a software component description, denoted $C_i$, is defined by the quadruplet: $C_i = <N, D, I^p, I^r>$;

Where:

– N is the name of the software component.
– D is the non-functional characteristics of the component which is independent from interfaces, such as the component version, the component technology, etc.
– $I^p$ is a set of provided interfaces.
– $I^r$ is a set of required interfaces.

• Each provided or required interface, denoted $I_j$, is defined by the couple $I_j = <N, Q>$;

Where N is the name of the interface and Q is the non- functional characteristics of the interface described according to the proposed QoS ontology.

- Given a developer request, where a set of required and provided interfaces is specified, the functional matching implies that each provided interface of the request is included in the corresponding provided interface of the candidate component whereas each required interface of this latter is included in the corresponding required interface of the request. Informally, this assumption means that the candidate component offers more functionalities (services) than the request and requires less external services than the request. Ideally, these interfaces are respectively similar.
- The QoS matching between two software components is performed after the functional matching that provides a set of candidate components that match, at the functional level, the request.

The QoS matching between software components consists of comparing their respective QoS profiles. A provided QoS profile describes a QoS profile related to a provided interface either specified for the request or for a candidate component. In addition, a required QoS profile corresponds to a required interface. Each request's QoS profile is compared to a component's QoS profile.

According to the proposed QoS ontology, a QoS profile is a conjunction of QoS metrics. Each QoS metric of the request's provided QoS profile must be matched against a QoS metric of the component's provided QoS profile. If there is a request's QoS metric that does not match against any QoS metric of the candidate component, the QoS matching fails. The matching between two required QoS profiles is computed following the same mechanism but with the order of the request and a candidate component reversed: Whereas the request's provided QoS profile is matched against the component's provided QoS profile, the required QoS profile of a candidate component is matched against the required QoS profile of the request.

### 4.1.1. Proposed Subsumption rules

The subsumption relationship between two provided or required QoS profiles depends on the relation between the QoS metric concepts. $R.I_j.Q$ denotes the QoS profile of the interface $I_j$ related to the request R and $C_i.I_j.Q$ denotes the QoS profile of the corresponding interface related to the candidate component $C_i$. $R.I_j.Q$ and $C_i.I_j.Q$ are described as follow:

$R.I_j.Q = M_1 \sqcap M_2 \sqcap \ldots \sqcap M_t$ where $M_{k\,/\,k=1,2,\ldots,t}$ are QoS metric concepts of the request R.

$C_i.I_j.Q = M'_1 \sqcap M'_2 \sqcap \ldots \sqcap M'_v$ where $M'_{k/k=1,2,\ldots,v}$ are QoS metric concepts of the candidate component $C_i$.

The matching between $R.I_j.Q$ and $C_i.I_j.Q$ is determined through the following rules:

- The matching between $R.I_j.Q$ and $C_i.I_j.Q$ is "Plugin" if and only if for each $M_k$, there is $M'_k$ such as $M'_k \sqsubseteq M_k$; we write $C_i.I_j.Q \sqsubseteq R.I_j.Q$;
- The matching between $R.I_j.Q$ and $C_i.I_j.Q$ is "Subsume" if and only if for each $M'_k$, there is $M_k$ such as $M_k \sqsubseteq M'_k$; we write $R.I_j.Q \sqsubseteq C_i.I_j.Q$;
- The match between $R.I_j.Q$ and $C_i.I_j.Q$ is "Exact" if and only if for each $M_k$, there is $M'_k$ such as $M_k \sqsubseteq M'_k$ and for each $M'_k$, there is $M_k$ such as $M'_k \sqsubseteq M_k$; we write, $R.I_j.Q \equiv C_i.I_j.Q$.

The "Plugin" match is the best relationship between two provided QoS profiles. The "Subsume" match is the best relationship between two required QoS profiles. The "Exact" match is the next best relationship between two QoS profiles either required or provided ones. Therefore, the "Fail" match is the worst result for QoS profiles matching.

### 4.1.2. QoS matching algorithm

The QoS matching between the request and a candidate component gives a vector, denoted QoSMatch. The QoSMatch is an array of integer elements and each entry in the vector is related to an interface and informs about the type of relationship between the associated QoS profiles. The length of the QoSMatch corresponds at most to the number of provided and required interfaces of a candidate component.

On the one hand, for a provided QoS profile of a candidate component, the matching result is detected and is organised in three levels as follows:

- "Plugin" is the best matching result. The corresponding element in the vector QoSMatch takes the value 2.
- "Exact" is the next best matching result. The corresponding element in the vector QoSMatch takes the integer value 1.
- "Fail" is the matching occurred otherwise.

On the other hand, for a required QoS profile of a candidate component, the "Subsume" match is considered at the top level of the matching results. The "Exact" match is the next best level. Otherwise the matching fails.

| Algorithme 1: QoS matching |
|---|
| *In the QoS matching algorithm, we use the following data constructs :* |
| R is the developper  request |
| $C_i$ are the candidate Components |
| *QoSMatch* is the QoS matching result between the request R and a candidate component $C_i$ |
| $(R.I_j.Q)$ represents the QoS profile of the interface $I_j$ of the request R |
| $(C_i.I_j.Q)$ represents the QoS profile of the interface $I_j$ of the candidate component $C_i$ |
| $\sum = \{(C_i,\ QoSMatch)\}$ is a set of couples. Each couple is composed of a candidate component and its matching vector with the request R |
| *The QoS matching algorithm is:* |
| 1. $\sum = \Phi$ |
| 2. *QoSMatch*= Array of integer |
| 3. for each $C_i$ ($1 \leq i \leq n$) do |
| 4.    begin |
| 5.        *QoSMatch.length* $\leftarrow$ *0*  *// initialise QoSMatch* |
| 6. for each $I_j$ ($1 \leq j \leq m$) do |
| 7.    begin |
| 8.       case match $(R.I_j.Q,\ C_i.I_j.Q)$ of |
| 9.          "*Plugin*": if $I_j \in I^p$ then *QoSMatch*[j]$\leftarrow$2 |
| 10.        "*Subsume*": if $I_j \in I^r$ then *QoSMatch*[j]$\leftarrow$2 |
| 11.        "*Exact*": *QoSMatch*[j]$\leftarrow$1 |
| 12.    end |

| | |
|---|---|
| 13. | if (*QoSMatch*.length>= µ ) then |
| | // *QoSMatch* contains at least µ elements |
| 14. | $\sum \leftarrow \sum \cup \{(C_i, QoSMatch)\}$ |
| 15. | end |
| 16. | return ($\sum$) |

Figure 4.  Proposed QoS matching algorithm

The idea of the QoS matching algorithm is the matching of all the request's QoS profiles against all the QoS profiles of the candidate component. For this, a "match" function (line 8) is performed and the concerned QoS profiles are passed as parameters. This function implements the subsumption rules described in section 4.1.1. The output of this algorithm is a set of candidate components in which at least µ interfaces (condition in line 13) match against the request. The value of µ is determined by the developer and reflects the developer constraints. The request can be relaxed by reducing the value of µ. All the candidate components that respond to the developer constraints are returned.

Table 1 shows an example (*c.f.* section 3.2) that serves to illustrate the QoS matching algorithm.

Table 1.  QoS descriptions of components.

| | R | $C_1$ | $C_2$ | $C_3$ |
|---|---|---|---|---|
| $I^p_1$ | FameOutput >=25 fps | Resolution >=72 dpi | FameRate >=30 fps | FameOutput >=25 fps |
| $I^p_2$ | StartUpTime <=10 ms | Response-Time<=10 ms | StartUpTime <=10 ms | TimeToRespond<=50 ms |
| $I^r_1$ | MTTF >=99.5% | MTTF >=99.5% | Reliability >=99.5% | Reliability >=99% |

Table 2 shows the matching results of the example of Table 1.

Table 2.  QoS matching results.

| $C_1$ | $C_2$ | $C_3$ |
|---|---|---|
| 1 | 1 | 1 |
| 1 | 1 | 1 |
| | 2 | 2 |

We note that the component $C_2$ and the component $C_3$ have matching values better than the matching values of the component $C_1$. The component $C_2$ and the component $C_3$ have the same QoSMatch vector but comparing their values they must have different ranks. Therefore, a component ranking process is needed and it is presented in the following section.

## 4.2. Component ranking process

The matching task is a crucial step in the component development process. However, an assessment task of the selected components has to be performed in order to choose the most appropriate components for selection.

In order to increase the result relevance, we compare the intervals of QoS values between the request and each candidate component. For this, we use the distance between attributes with interval values. The ranking process is often separated from the matching one because the ontological reasoning is a computationally expensive task. However, the component ranking process takes into account the QoS matching results.

### 4.2.1. QoS normalisation

The normalisation is an important task in the analysis step to compare data having different domain values. The normalisation makes data look increasingly similar. An attribute is normalised by scaling its values so that they fall within a small-specified range, such as 0 to 1.

In our work, we use the min-max normalisation [20] which preserves the relationships among the original QoS values. Given a QoS value $q_i \in$ [min, max], it is mapped into normalised QoS value $q'_i$ by using the following formula:

$$q'_i = \frac{q_i - \min}{\max - \min}$$

### 4.2.2. Semantic dissimilarity measure

Our idea is to sort the matched components according to a new semantic dissimilarity measure, denoted CRank. The CRank measure takes into account the values of QoS metrics and is calculated as follow:

$$CRank = \sum_{j=1}^{QoSMatch.length} \frac{1}{QoSMatch[j]} \sum_{K=1}^{\min(t,v)} \delta_{(M_k , M'_k)}$$

Where QoSMatch is the QoS matching vector (c.f. section 4.1.2) and $\delta_{(M_k , M'_k)}$ is the distance between the intervals

$$[M_k . \min Value, M_k . \max Value] \quad \text{and} \quad [M'_k . \min Value, M'_k . \max Value]$$

$\delta_{(M_k , M'_k)}$ is calculated by the following formula:

$$\delta_{(M_k , M'_k)} = \frac{\left| M_k . \max Value - M'_k . \max Value \right| + \left| M_k . \min Value - M'_k . \min Value \right|}{2}$$

Where

- $M_k . \min Value$ and $M_k . \max Value$ are, respectively, the minimum and the maximum normalised values related to a QoS metric specified in a request's QoS profile (R.I$_j$.Q);

25

- $M'_k . \min Value$ and $M'_k . \max Value$ are, respectively, the minimum and the maximum normalised values related to a QoS metric specified in a component's QoS profile $(C_i.I_j.Q)$.

- The QoS metrics considered are those common to the QoS profiles $R.I_j.Q$ and $C_i.I_j.Q$.

*Example:*

Based on the example of Table1, the ranking values of the components $C_2$ and $C_3$ are presented in Table 3.

Table 3. Component ranking results.

| CMatch | R | $C_2$ | $C_3$ |
|---|---|---|---|
| 1 | [0, 1] | [0.11, 1] | [0, 1] |
| 1 | [0, 0] | [0, 0] | [0, 0.04] |
| 2 | [0.5, 1] | [0.5, 1] | [0, 1] |
| **CRank** | | **0.055** | **0.145** |

The best component is the component $C_2$ which has the smallest dissimilarity measure.

### 4.2.3. Component ranking algorithm

| Algorithm 2: Component ranking |
|---|
| *In the component ranking algorithm, we use the following data constructs :*<br><br>R is the developper request<br><br>$C_i$ are the candidate Components<br><br>*QoSMatch* is the QoS matching result between the request R and one candidate component $C_i$<br><br>$(R.I_j.Q)$ represents the QoS profile of the interface $I_j$ of the request R<br><br>$(C_i.I_j.Q)$ represents the QoS profile of the interface $I_j$ of the component $C_i$<br><br>$\delta_{(M_k, M'_k)}$ is the distance between two intervals of values related to the QoS metrics $M_k$ and $M'_k$<br><br>$\sum' = \{(C_i, CRank)\}$ is a set of the candidate components with their ranks<br><br>*The component ranking algorithm is:*<br><br>1. $\sum' = \Phi$<br><br>2. *CRank* $\leftarrow 0$<br><br>3. for each $C_i$ (1≤i≤n) do<br><br>4.     for each $I_j$ (1≤j≤m) do<br><br>5.         begin<br><br>6.             $\delta_{(M_k, M'_k)} \leftarrow 0$<br><br>7.             for each $(M_k, M'_k)$ (1≤k≤min(t,v)) |

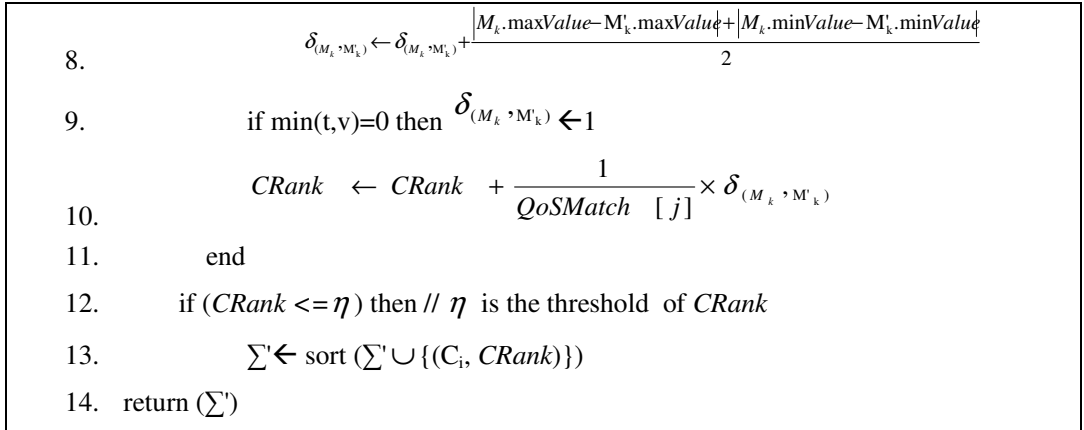

8.     $\delta_{(M_k,\,'M'_k)} \leftarrow \delta_{(M_k,\,'M'_k)} + \dfrac{|M_k.maxValue - M'_k.maxValue| + |M_k.minValue - M'_k.minValue|}{2}$

9.     if min(t,v)=0 then $\delta_{(M_k,\,'M'_k)} \leftarrow 1$

10.     $CRank \leftarrow CRank + \dfrac{1}{QoSMatch\;[\,j\,]} \times \delta_{(M_k,\,'M'_k)}$

11.     end

12.     if ($CRank <= \eta$ ) then // $\eta$ is the threshold of $CRank$

13.     $\sum' \leftarrow$ sort ($\sum' \cup \{(C_i, CRank)\}$)

14.   return ($\sum'$)

Figure 5. Proposed component ranking algorithm

The idea of the component ranking algorithm is to calculate a rank for each component selected during the QoS matching process. The ranking algorithm takes into account the QoS metrics values and combines them with the matching result. For each QoS profile, the intervals of metrics values are compared and divided by the corresponding value of QoSMatch (line 10). Finally, the set of candidate components, whose ranks below a threshold (line 12), is sorted (line 13) then returned (line 14).

## 5. EXPERIMENT

In order to measure the performance and the utility of our approach, we experimented it on a set of candidate components and a set of requests. Thus, we worked on video camera components from the multimedia domain. Nevertheless, we are only focused on the subsumption relation between the QoS metrics Reliability and MTTF (Mean Time To Failure) such as: MTTF ⊑ Reliability.

Our experiment is applied, with 8 requests (*cf.* Table 4), on the set of candidate components. For each request, the relevance of components determined by the developer is used to calculate the precision and the recall of the selection process. The precision is the number of relevant components compared to the total number of selected components. Whereas, the recall is the number of relevant selected components compared to the number of relevant components in the set of candidate components.

Table 4. The set of requests.

| Request | Request description |
| --- | --- |
| $R_1$ | Reliability>=99.5 |
| $R_2$ | ResponseTime<=20 |
| $R_3$ | FrameRate>=30 |
| $R_4$ | MTTF>=99, ResponseTime<=50 |
| $R_5$ | MTTF>=99.5, 25<=FrameRate<=30 |
| $R_6$ | ResponseTime<=10, FrameRate=30 |
| $R_7$ | Reliability>=99, ResponseTime <=50, 60<=FrameRate<=72 |
| $R_8$ | MTTF>=99, ResponseTime<=50, 60<=FrameRate<=72 |

In a first step, we applied the QoS matching mechanism for selecting components. In a second step, the experiment is extended to the component ranking mechanism. The performance of the selection mechanism is measured by both the recall and the precision. The results indicate that the performance of the component selection is good when using only the subsumption reasoning but it is better after using the proposed semantic dissimilarity measure (cf. section 4.2.2).

The algorithm of matching has both a good precision and a good recall. When applying it on the set of software components, it achieves a precision of 0.604 and a recall of 0.937.
The table 5 shows the average of the recall and the precision corresponding to the eight requests with the set of candidate components.

Table 5. Precision and Recall after matching.

| Request | Precision | Recall |
|---------|-----------|--------|
| $R_1$ | 0.2 | 0.5 |
| $R_2$ | 1 | 1 |
| $R_3$ | 1 | 1 |
| $R_4$ | 0.6 | 1 |
| $R_5$ | 0.2 | 1 |
| $R_6$ | 1 | 1 |
| $R_7$ | 0.5 | 1 |
| $R_8$ | 0.33 | 1 |
| **Average** | **0.604** | **0.937** |

The ranking algorithm allows the usage of a semantic dissimilarity measure. The experiment is done with the same sets of requests and candidate components. The results indicate that the precision is increased by 0.239 and the recall is decreased by 0.062. The new results are shown in Table 6.

Table 6. Precision and Recall after matching.

| Request | Precision | Recall |
|---------|-----------|--------|
| $R_1$ | 1 | 0.5 |
| $R_1$ | 1 | 1 |
| $R_1$ | 1 | 0.5 |
| $R_1$ | 0.75 | 1 |
| $R_1$ | 0.5 | 1 |
| $R_1$ | 1 | 1 |
| $R_1$ | 1 | 1 |
| $R_1$ | 0.5 | 1 |
| **Average** | **0.843** | **0.875** |

Comparing the previous results, we conclude that the performance of the selection process is better when performing both the matching and the ranking mechanisms.

# 6. CONCLUSION

The Component Selection problem has been addressed in various researches. The most solutions consider only the functional requirements. In CBSE field, Few works deal with the QoS requirements because of the complexity of this task. In this work, the heterogeneous concepts of QoS are well-defined through an ontological formalization which improves the understanding of QoS metrics and values and so, these concepts can be machine-readable.

The integration of the QoS metrics and their values in the selection process is proposed. We focus on the benefits of the ontology reasoning and we propose two algorithms. The QoS matching algorithm, which is based on ontology reasoning, gives good results for component selection. In addition, the proposed component ranking improves the performance of the selection process.

Currently, we are experimenting the relevance of the proposed semantic dissimilarity measure with greater sets of requests and software components.